# Change in stacking order and Lifshitz transition in Bi-layer Graphene

Partha Goswami

Physics Department, D.B.College, University of Delhi, Kalkaji, New Delhi-110019, India
*Email id of communicating author:*physicsgoswami@gmail.com*

**Abstract** We consider the AB-(Bernal) stacking for the bi-layer graphene (BLG) system and assume that a perpendicular electric field is created by the external gates deposited on the BLG surface. In the basis $(A_1, B_2, A_2, B_1)$ for the valley **K** and the basis $(B_2, A_1, B_1, A_2)$ for the valley **K′**, we show the occurrence of trigonal warping [ see A. S. Núñez et al.,arXiv:1012.4318] , that is, splitting of the energy bands or the density of states on the $k_x$ - $k_y$ plane into four pockets comprising of the central part and three legs due to a (skew) interlayer hopping between $A_1$ and $B_2$. The hopping between $A_1 - B_2$ leads to a concurrent velocity $v_3$ in addition to the Fermi velocity $v_F$. Our noteworthy outcome is that the above-mentioned topological change, referred to as the Lifshitz transition [I. M. Lishitz, Zh. Exp. Teor. Fiz., 38, 1565 (1960) (Sov. Phys. JETP 11, 1130 (1960));Y. Lemonik, I.L. Aleiner, C. Toke, and V.I. Fal'ko; arXiv:1006.1399], is entirely bias-tunable. Furthermore, the many-body effects, which is known to yield logarithmic renormalizations [C. Hwang, D. A. Siegel, Sung-Kwan Mo, W. Regan, A. Ismach, Y. Zhang, A. Zettl and A. Lanzara, arXiv:1208.0567] in the band dispersions of monolayer graphene, is found to have significant effect on the bias-tunability of this transition. We also consider the system where the A atoms of the two layers are over each other and the B atoms of the layers are displaced with respect to each other. The trigonal warping is found to be absent in this case. Instead, the Fermi energy density of states for zero bias corresponds to the inverted sombrero-like structure. The structure is found to get deformed due to the increase in the bias.

**Keywords:** AB-(Bernal) stacking, trigonal warping, Lifshitz transition, Logarithmic renormalizations, Inverted sombrero.

**PACS:** 73.22.-f.

## Main text

In two very exhaustive review articles Castro Neto et al. **[1]** have discussed many peculiar properties of graphene. These peculiarities have greatly intrigued physicists in recent years. In the monolayer graphene (MLG), the charge carriers are mass-less Dirac particles of chiral nature near neutrality points. The bi-layer graphene(BLG) presents an entirely different landscape where the two layers are coupled by weak van der Waals forces. The carriers, for example, in the (Bernal AB-stacked) bi-layer graphene (BLG) are neither Dirac nor Schrodinger fermions. In the Bernal stacking the two layers in the bi-layer graphene, consisting of two coupled honeycomb lattices with basis atoms $(A_1, B_1)$ and $(A_2, B_2)$ in the bottom and the top layers, respectively, are arranged in $(A_2, B_1)$ fashion. That is, the A-carbon of the upper sheet lies on top of the B-carbon of the lower one. The intra-layer coupling between $A_1$ and $B_1$ and $A_2$ and $B_2$ is $\gamma_0 = 3.16$ eV. The strongest interlayer coupling is between $A_2$ and $B_1$ with coupling constant $\gamma_1 = 0.39$ eV. We consider a (skew) interlayer hopping between $A_1$ and $B_2$ with strength $\gamma_3 = 0.315$ eV. This introduces an additional velocity $v_3 = (3/2)a\gamma_3/\hbar = 5.9 \times 10^4$ m-s$^{-1}$ and causes a significant trigonal warping **[2]** of the energy dispersion. The values of these hopping integrals will be taken to be same as in Ref. **[3]** in our calculation below. The lattice model in real space, for the BLG system, can be written in the tight-binding form with an electrostatic bias V as

$$H = \sum_{i,\sigma} V\{a^\dagger_{1,i,\sigma} a_{1,i,\sigma} + b^\dagger_{1,i,\sigma} b_{1,i,\sigma} - a^\dagger_{2,i,\sigma} a_{2,i,\sigma}$$

$$- b^\dagger_{2,i,\sigma} b_{2,i,\sigma}\} - \gamma_0 \sum_{\langle i,j \rangle,m,\sigma} (a^\dagger_{m,i,\sigma} b_{m,j,\sigma} + b^\dagger_{m,j,\sigma} a_{m,i,\sigma})$$

$$- \gamma_1 \sum_{i,\sigma} (a^\dagger_{2,i,\sigma} b_{1,i,\sigma} + b^\dagger_{1,i,\sigma} a_{2,i,\sigma})$$

$$- \gamma_3 \sum_{\langle i,j \rangle,\sigma} (a^\dagger_{1,i,\sigma} b_{2,j,\sigma} + b^\dagger_{2,j,\sigma} a_{1,i,\sigma}) \qquad (1)$$

where the NN hopping integral corresponds to the index $\langle i,j \rangle$ in the third term in (1). The operators $a^\dagger_{m,j,\sigma}$ and $b^\dagger_{m,j,\sigma}$ with spin σ, respectively, correspond to the fermion creation operators for A and B sub-lattices in the m = 1,2 layer. Close to the Dirac point in the Brillouin zone, upon expanding the momentum, the low-energy Hamiltonian for the Bernal AB-stacked BLG could be written in a compact form $H = \sum_{\delta\mathbf{k}} \Psi^\dagger_{\delta\mathbf{k}} H(\delta\mathbf{k}) \Psi_{\delta\mathbf{k}}$ in the basis $(A_1, B_2, A_2, B_1)$ in the valley **K**. The row vector $\Psi^\dagger_{\delta\mathbf{k}} = (a^\dagger_1(\delta\mathbf{k})\ b^\dagger_2(\delta\mathbf{k})\ a^\dagger_2(\delta\mathbf{k})\ b^\dagger_1(\delta\mathbf{k}))$ ; $a^\dagger_1(\delta\mathbf{k}), b^\dagger_2(\delta\mathbf{k})$, etc. stand for the fermion creation operators in the momentum space. For the valley **K′**, the appropriate basis is $(B_2, A_1, B_1, A_2)$. We assume that a perpendicular electric field is (electrostatic bias V) created by the external gates deposited on the BLG surface. This induces a gap in the energy spectrum through a charge imbalance between the two graphene layers. The Hamiltonian matrix $H(\delta\mathbf{k})$ is given by

$$H(\delta\mathbf{k}) = \xi \begin{pmatrix} V & v_3\delta\mathbf{k} & 0 & v_F\delta\mathbf{k}^* \\ v_3\delta\mathbf{k}^* & -V & v_F\delta\mathbf{k} & 0 \\ 0 & v_F\delta\mathbf{k}^* & -V & \xi\gamma_1 \\ v_F\delta\mathbf{k} & 0 & \xi\gamma_1 & V \end{pmatrix}, (2)$$

where $v_F$ is Fermi velocity (the speed of electrons in the vicinity of a Dirac point in the absence of interlayer hopping and is equal to $8 \times 10^5$ m-s$^{-1}$), $\delta\mathbf{k} = (\delta k_x + i\,\delta k_y)$ is a complex number and $\xi = \pm 1$; $\xi = +1$ corresponds to the valley **K** and $\xi = -1$ to the valley **K′**. We shall now consider the many-body effects only on the dominant terms $(v_F\delta\mathbf{k}, v_F\delta\mathbf{k}^*)$ above. A similar exercise for all the terms have been carried out by C. T˝oke and V. I. Fal'ko**[4]** in theHartree-Fock approximation. We feel that a recently reported crucial many-body effect**[5]** in the band dispersions of monolayer graphene needs to be included in a description of the bi-layer system. In other previous approaches [6,7] for BLG, all effects of Coulomb interactions are ignored except the Coulomb interaction for an electron and hole adjacent to each other but in opposite layers. It may be noted that the path integral approach requires no single-particle approximation and therefore many-body effects emerge naturally. Since we shall not adopt this rigorous formalism in the present letter, our approach is essentially an approximation requiring the introduction of the many-body effects by using the Dyson's equation.

For the purpose stated above, one may write few unperturbed thermal averages determined by the Hamiltonian in (2),viz. $G^0_{AA,m}(\delta \mathbf{k},\tau) = -\langle T\{a_{m,\delta \mathbf{k}}(\tau) a^\dagger_{m,\delta \mathbf{k}}(0)\}\rangle$, $G^0_{AB,m}(\delta \mathbf{k},\tau) = -\langle T\{a_{m,\delta \mathbf{k}}(\tau) b^\dagger_{m,\delta \mathbf{k}}(0)\}\rangle$, $G^0_{BA,m}(\delta \mathbf{k},\tau) = -\langle T\{b_{m,\delta \mathbf{k}}(\tau) a^\dagger_{m,\delta \mathbf{k}}(0)\}\rangle$, and $G^0_{BB,m}(\delta \mathbf{k},\tau) = -\langle T\{b_{m,\delta \mathbf{k}}(\tau) b^\dagger_{m,\delta \mathbf{k}}(0)\}\rangle$ with m = 1,2. Here T is the time-ordering operator which arranges other operators from right to left in the ascending order of imaginary time τ. The Fourier coefficients of these temperature functions are $G_{\alpha\beta,m}(\mathbf{k},\omega_n) = \int_0^\beta d\tau\, e^{i\omega_n\tau} G_{\alpha\beta,m}(\mathbf{k},\tau)$ (where the Matsubara frequencies are $\omega_n = [(2n+1)\pi/\beta]$ with n = 0,±1,±2,…. and $\beta = (k_B T)^{-1}$). We obtain for the $m^{th}$ sheet $G^0_{AA,m}(\delta \mathbf{k},\omega_n) = G^0_{BB,m}(\delta \mathbf{k},\omega_n) \approx (1/2)[(i\omega_n - E_+(\delta \mathbf{k}))^{-1} + (i\omega_n - E_-(\delta \mathbf{k}))^{-1}]$, and so on. In Eq.(2), upon retaining only the terms $(v_F \delta \mathbf{k}, v_F \delta \mathbf{k}^*)$, we obtain $E_\pm(\delta \mathbf{k}) = \pm \hbar v_F |\delta \mathbf{k}|$. It was proposed by Castro Neto et al. [1,5] that, unlike the linear real self-energy of a Fermi liquid, when monolayer graphene(MLG) is near the charge neutrality point the electron-electron interaction leads to a self-energy involving logarithmic term given by $\sum'(k) = (\alpha \hbar v_0/4)(k-k_F) \ln(k_c/(k-k_F))$. This is the 'so called' marginal Fermi liquid self-energy function for MLG. Here $k_F = 1.703$ A$^{o\ -1}$ is the Fermi wave-number along the Γ- K direction, $v_0 = 0.85 \times 10^6$ m/s is the Fermi velocity for the dielectric constant ε = 6.4 ± 0.1, $k_c$ is the momentum cut-off~ $k_F$, and α = 0.40 ± 0.01 is a dimensionless fine-structure constant (or the strength of electron-electron interactions) defined as $(e^2/4\pi\varepsilon\hbar v_0)$. In terms of the logarithmic self-energy, using the Dyson's equation, a full propagator for the $m^{th}$ sheet $G_{\alpha\beta,m}(\delta \mathbf{k},\omega_n)$ could be approximated as $G^0_{\alpha\beta,m}(\delta \mathbf{k},\omega_n)/(1-2G^0_{\alpha\beta,m}(\delta \mathbf{k},\omega_n)\sum(k))$, where $\sum(k) = (\alpha\hbar v_0/8)(k-k_F) \ln(k_c/(k-k_F))$. The approximate analytic form of the full propagator is

$(1/2)[1+(\sum(k)/\sqrt{\{(\hbar v_F |\delta \mathbf{k}|)^2 + \sum^2(k)\}})] \times [i\omega_n - \varepsilon'_1(k)]^{-1}$

$+(1/2)[1-(\sum(k)/\sqrt{\{(\hbar v_F |\delta \mathbf{k}|)^2 + \sum^2(k)\}})] \times [i\omega_n - \varepsilon'_2(k)]^{-1}$,

$\varepsilon'_1(\delta \mathbf{k}) = \sqrt{\{(\hbar v_F |\delta \mathbf{k}|)^2 + \sum^2(k)\}} + \sum(\delta \mathbf{k})$,

$\varepsilon'_2(\delta \mathbf{k}) = -\sqrt{\{(\hbar v_F |\delta \mathbf{k}|)^2 + \sum^2(k)\}} + \sum(\delta \mathbf{k})$.   (3)

The poles $(\varepsilon'_1(k),\varepsilon'_2(k))$ allow us to re-construct the intra-layer coupling between $A_1$ and $B_1$ and $A_2$ and $B_2$; the interlayer coupling between $A_2$ and $B_1$ (with coupling constant $\gamma_1$) and the (skew) interlayer hopping between $A_1$ and $B_2$ (with strength $\gamma_3$) remains unaffected by the reconstruction as stated above. Effectively, we have assumed here that the inter-layer separation is larger than the intra-layer nearest neighbor separation. A plot of the ratio Re($\sum'(k))/\gamma_1$ as a function of momentum is shown in Figure 1. Close to the Dirac points **K** $(2\pi/3a, 2\pi/3\sqrt{3}a)$ and **K'** $(2\pi/3a, -2\pi/3\sqrt{3}a)$, where ka = 2.4184, we find that the self-energy corrections are very significant as these may be greater than the linear terms in momentum in $(\varepsilon'_1(k),\varepsilon'_2(k))$.

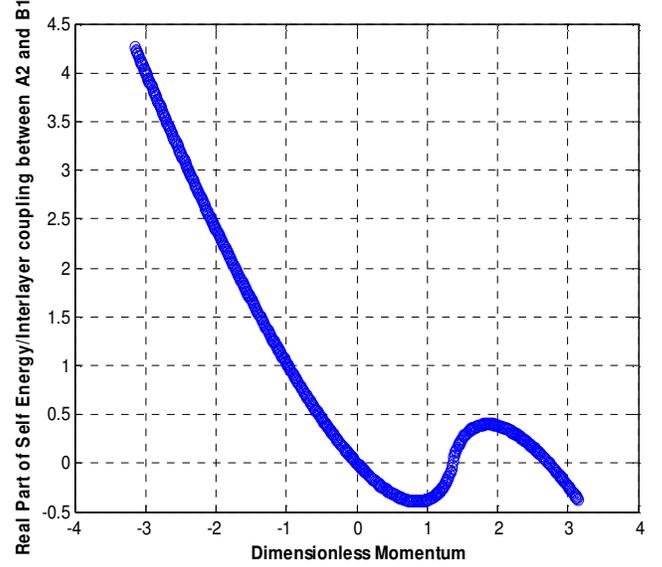

**Figure 1.** A plot of the ratio Re($\sum'(k))/\gamma_1$ as a function of momentum. The numerical values used in the plot could be found in the text.

With the self-energy correction, the matrix in (2) may be re-written as

$$H(\delta \mathbf{k}) \approx \xi \begin{pmatrix} V & v_3 \delta \mathbf{k} & 0 & \varepsilon^*_1(\delta \mathbf{k}) \\ v_3 \delta \mathbf{k}^* & -V & \varepsilon_2(\delta \mathbf{k}) & 0 \\ 0 & \varepsilon^*_2(\delta \mathbf{k}) & -V & \xi\gamma_1 \\ \varepsilon_1(\delta \mathbf{k}) & 0 & \xi\gamma_1 & V \end{pmatrix}, \quad (4)$$

where $\varepsilon_1(\delta \mathbf{k}) = \sqrt{\{(\hbar v_F \delta \mathbf{k})^2 + \sum^2(\delta \mathbf{k})\}} + \sum(\delta \mathbf{k})$, and $\varepsilon_2(\delta \mathbf{k}) = -\sqrt{\{(\hbar v_F \delta \mathbf{k})^2 + \sum^2(\delta \mathbf{k})\}} + \sum(\delta \mathbf{k})$. The eigenvalue (λ) equation of the matrix in (4) is a quartic:

$\lambda^4 - 2\lambda^2 \{|\varepsilon_1|^2 + |\varepsilon_2|^2 + V^2 + (\gamma_1^2/2) + v_3^2 |\delta \mathbf{k}|^2\} - 2\lambda V\{|\varepsilon_1|^2 - |\varepsilon_2|^2\}$

$+ [V^4 + V^2 \gamma_1^2 - V^2(|\varepsilon_1|^2 + |\varepsilon_2|^2) + v_3^2 |\delta \mathbf{k}|^2 (V^2 + \gamma_1^2)$

$-\xi(\varepsilon_1 \varepsilon_2 \delta \mathbf{k} + \varepsilon_1^* \varepsilon_2^* \delta \mathbf{k}^*) v_3 \gamma_1 + |\varepsilon_1|^2 |\varepsilon_2|^2] = 0.$ (5)

If one ignores the self-energy correction altogether, Eq.(5) reduces to a bi-quadratic whose solutions are easy to obtain. We obtain four bands, $E_p^\pm(\delta \mathbf{k})$, p = 1,2, as reported by Fal'ko et al.[2] with

$E_p(\delta \mathbf{k})^2 = [\mathcal{E}_1(\delta \mathbf{k})^2 + V^2]$

$$+(-1)^p \sqrt{[\mathcal{E}_1(\delta\mathbf{k})^4 + 4(v_F|\delta\mathbf{k}|)^2 V^2 - \mathcal{E}_2(\delta\mathbf{k})^4]}, \quad (6)$$

where $E_1$ and $E_2$, respectively, describes the lower and higher energy bands, and

$$\mathcal{E}_1(\delta\mathbf{k})^2 = (v_F|\delta\mathbf{k}|)^2 + (1/2)(\gamma_1^2 + (v_3|\delta\mathbf{k}|)^2),$$

$$\mathcal{E}_2(\delta\mathbf{k})^4 = (v_F|\delta\mathbf{k}|)^4 + (v_3|\delta\mathbf{k}|)^2 \gamma_1^2$$
$$- 2 v_3 v_F^2 \xi \gamma_1 |\delta\mathbf{k}|^3 \cos(3\varphi). \quad (7)$$

We have parameterized $\delta\mathbf{k}$ writing $\delta k_x = |\delta\mathbf{k}|\cos(\varphi)$ and $\delta k_y = |\delta\mathbf{k}|\sin(\varphi)$ which gives $\delta k = |\delta\mathbf{k}|\exp(i\varphi)$. The effect of valley state plus the the skew interlayer hopping between $A_1 - B_2$, given by the last term in $\mathcal{E}_2(\delta\mathbf{k})^4$, on the four bands are found to be extremely sensitive to the bias. The bands $E_p(\delta\mathbf{k})$ splits into four pockets comprising of the central part and three legs[2] for $\varphi = \{0, 2\pi/3, 4\pi/3\}, \{\pi/3, \pi, 5\pi/3\}$. We note that such splitting is an indication of the Lifshitz transition[8]. We have shown in Figure 2 this topological change in the Fermi surface density of states(DOS). We have started with the electrostatic bias $(V/\gamma_1) = 0.1$ at which the change sets in. The plots in Figure 2(a) and 2(b) correspond to $(V/\gamma_1) = 0.107$. A higher value of $(V/\gamma_1)$, as much as 0.17, almost obliterates the four-pocket feature from the DOS. Thus, the transition appears to be bias-tunable.

We obtain the solutions of Eq.(5) using the Ferrari's method of solving a quartic. Given the general quartic $Ax^4 + Bx^3 + Cx^2 + Dx + E = 0$, its solution could be found by means of the following algebra: We introduce $\alpha = -(3B^2/8A^2) + (C/A)$, $\beta = (B^3/8A^3) - (BC/2A^2) + (D/A)$, and $\gamma = -(3B^4/256A^4) + (CB^2/16A^3) - (BD/4A^2) + (E/A)$. In the present problem, $A = 1$, $B = 0$, so $\alpha = C$, $\beta = D$, and $\gamma = E$. We further define $P = -(\alpha^2/12) - \gamma$, $Q = -(\alpha^3/108) + (\alpha\gamma/3) - (\beta^2/8)$, and $R = -(Q/2) \pm \sqrt{\{(Q^2/4) + (P^3/27)\}}$. This ultimately yields the single-particle excitation spectra $E^{(r=(I,II),s)}(\delta\mathbf{k})$ given by $E^{(r=(I,II),s)}(\delta\mathbf{k}) = [rW(\delta\mathbf{k})/2 + s(1/2)\sqrt{\{-(3\alpha + 2\breve{Y} + (2r\beta/W(\delta\mathbf{k})))\}}]$, where $W = \sqrt{(\alpha + 2y)}$, $\breve{Y} = -(5\alpha/6) + U - (P/3U)$, $U = \sqrt[3]{R}$, $r$ is equal to $(\pm 1)$ with $r = +1$ corresponding to the branch (I) and $r = -1$ to the branch (II) and for a given r we have $s = \mp 1$. The single-particle spectral function or density of states(DOS) is given by a retarded Green's function. We find that the DOS is given by a sum of four $\delta$ functions at the quasi-particle energies. We have plotted in momentum space(see Figure 3) the Fermi surface DOS with these bands and an artificial level broadening $(\acute{\Gamma}/\gamma_1) = 0.0001$ once again. We have assumed $(v_F/a\gamma_1) = 7.9$. The remaining numerical values are $(V/\gamma_1) = 0.20$ and $(v_3/a\gamma_1) = 0.7949$. We find that in this case the Lifshitz transition[8] sets in at $(V/\gamma_1) \sim 0.17$ and a higher value of $(V/\gamma_1)$, as much as 0.22, almost obliterates the four-pocket feature from the DOS. We, thus, find that in the presence of many-body effects higher bias is required for the occurrence of the transition.

In this letter we also consider the system where the A atoms of the two layers are over each other and the B atoms of the layers are displaced with respect to each other. In the tight-binding description [1], the Hamiltonian with the electrostatic bias (V) is given by

$$H = \sum_{i,\sigma} V\{(a^\dagger_{1,i,\sigma} a_{1,i,\sigma} + b^\dagger_{1,i,\sigma} b_{1,i,\sigma})$$
$$-(a^\dagger_{2,i,\sigma} a_{2,i,\sigma} + b^\dagger_{2,i,\sigma} b_{2,i,\sigma})\} - \gamma_0 \sum_{\langle i,j \rangle, m, \sigma} (a^\dagger_{m,i,\sigma} b_{m,j,\sigma} + b^\dagger_{m,j,\sigma} a_{m,i,\sigma}) - \gamma_1 \sum_{j,\sigma} (a^\dagger_{1,j,\sigma} a_{2,j,\sigma} + a^\dagger_{2,j,\sigma} a_{1,j,\sigma}) - \gamma_3 \sum_{\langle i,j \rangle, \sigma} (a^\dagger_{1,i,\sigma} b_{2,j,\sigma} + a^\dagger_{2,i,\sigma} b_{1,j,\sigma} + b^\dagger_{2,j,\sigma} a_{1,i,\sigma}$$
$$+ b^\dagger_{1,j,\sigma} a_{2,i,\sigma}) - \gamma_4 \sum_{\langle i,j \rangle, \sigma} (b^\dagger_{1,i,\sigma} b_{2,j,\sigma} + b^\dagger_{2,j,\sigma} b_{1,i,\sigma}). \quad (8)$$

As before, the intra-layer coupling between $A_1$ and $B_1$ and $A_2$ and $B_2$ is $\gamma_0 = 3.16$ eV. The strongest direct interlayer coupling is between $A_1$ and $A_2$ with coupling constant $\gamma_1 = 0.39$ eV. The skew interlayer hopping between $A_1$ and $B_2$ (and between $A_2$ and $B_1$) with strength $\gamma_3 = 0.315$ eV introduces an additional velocity $v_3 = (3/2)a\gamma_3/\hbar = 5.9 \times 10^4$m·s$^{-1}$. These numerical values are almost the same as in ref.[3]. Close to the Dirac point $\mathbf{K}$ in the Brillouin zone, upon expanding the momentum, this Hamiltonian with the electrostatic bias could be written in the compact form $H = \sum_{\delta\mathbf{k}} \Psi^\dagger_{\delta\mathbf{k}} H(\delta\mathbf{k}) \Psi_{\delta\mathbf{k}}$ in the basis $(A_1, B_1, A_2, B_2)$ where $\Psi^\dagger_{\delta\mathbf{k}} = (a^\dagger_1(\delta\mathbf{k}) \; b^\dagger_1(\delta\mathbf{k}) \; a^\dagger_2(\delta\mathbf{k}) \; b^\dagger_2(\delta\mathbf{k}))$,

$$H(\delta\mathbf{k}) = \begin{pmatrix} V & v_F\delta\mathbf{k} & \gamma_1 & v_3\delta\mathbf{k}^* \\ v_F\delta\mathbf{k}^* & V & v_3\delta\mathbf{k} & 0 \\ \gamma_1 & v_3\delta\mathbf{k}^* & -V & v_F\delta\mathbf{k} \\ v_3\delta\mathbf{k} & 0 & v_F\delta\mathbf{k}^* & -V \end{pmatrix}. \quad (9)$$

As before, $\delta k = \delta k_x + i\delta k_y$ is a complex number. For the valley $\mathbf{K}'$, the basis would be $(B_1, A_1, B_2, A_2)$. In writing Eq.(9) we have ignored $\gamma_4$ term as this term is smaller than the others. As before, the density of states(DOS) is given by a retarded Green's function. We find that the DOS is given by a sum of four $\delta$ functions at the quasi-particle energies. We have also contour plotted the DOS (Figure 4) with an artificial level broadening $(\acute{\Gamma}/\gamma_1) = 0.0001$. The figures (a), and (b), respectively, correspond to the $(V/\gamma_1) = 0$, and $(V/\gamma_1) = 0.17$. We notice a bias induced change in the topology of the Fermi surface DOS in this case as well. However, this transition is quite different from the earlier one as the familiar trigonal warping is absent. We find that without bias the system is more stable compared to the case where the bias is present as there is slight increase in the free energy in the latter case. However, the activation energy necessary to destroy this AA-stacked phase is over one order of magnitude less than that of the AB-stacked system to make it meta-stable at room temperature.

In conclusion, this striking reconstruction of the Fermi surface at low densities may lead to an asymmetry in the conductivity under electron or hole doping. We have, however, only estimated the decrease in the electronic specific heat due to this bias-tunable transition. It is found to be close to 10%. Thus, the Lifshitz transition, in principle, is detectable in the heat capacity measurements. It must be added that the experimental observation of the decrease is quite a difficult proposition, for the dominant phononic contribution is expected to overshadow the anomaly in the measurements.

.

**FIGURES**

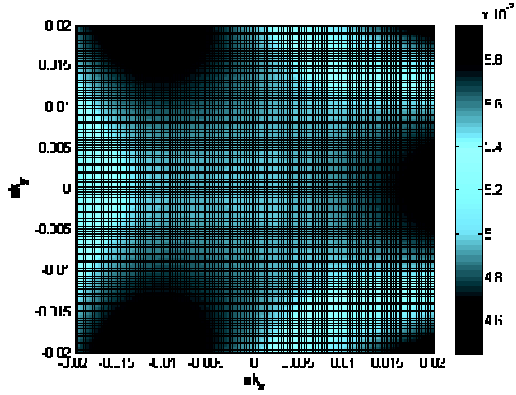

(a)

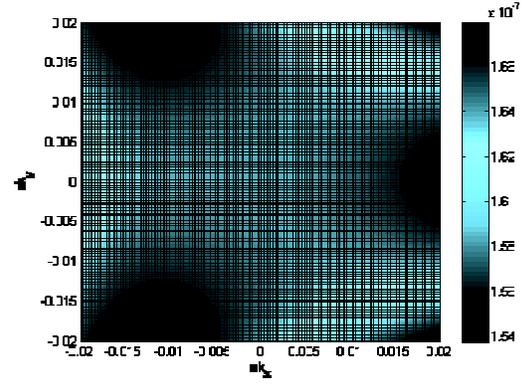

(a)

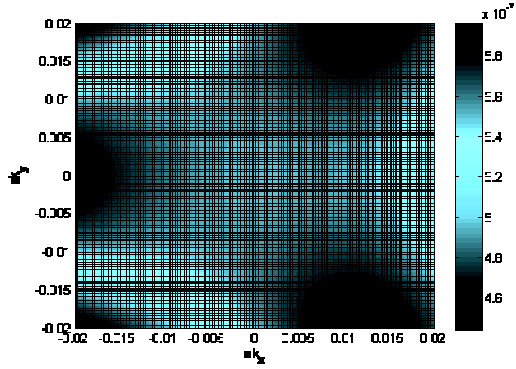

(b)

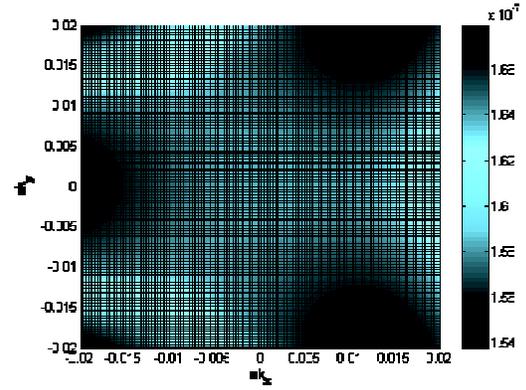

(b)

**Figure 2.** We have contour plotted the Fermi surface DOS obtainable from Eq.(6) in the momentum space with an artificial level broadening $(\acute{\Gamma}/\gamma_1) = 0.0001$. We find that the DOS is given by a sum of four δ functions at the quasi-particle energies. We notice that the trigonal warping splits the surface into four pockets comprising of the central part and three legs both for the valley states **K** (Figure(a)) and **K′** (Figure(b)). We have assumed $(v_F/a\gamma_1) = 7.9$. The remaining numerical values are $(V/\gamma_1) = 0.1$ and $(v_3/a\gamma_1) = 0.7949$.

**Figure 3.** We have contour plotted the Fermi surface DOS obtainable from Eq.(5) including the many-body effect in the momentum space with an artificial level broadening $(\acute{\Gamma}/\gamma_1) = 0.0001$. The trigonal warping splits the surface into four pockets comprising of the central part and three legs both for the valley states **K** (Figure(a)) and **K′** (Figure(b)) as before. We have assumed $(v_F/a\gamma_1) = 7.9$. The remaining numerical values are $(V/\gamma_1) = 0.20$ and $(v_3/a\gamma_1) = 0.7949$.

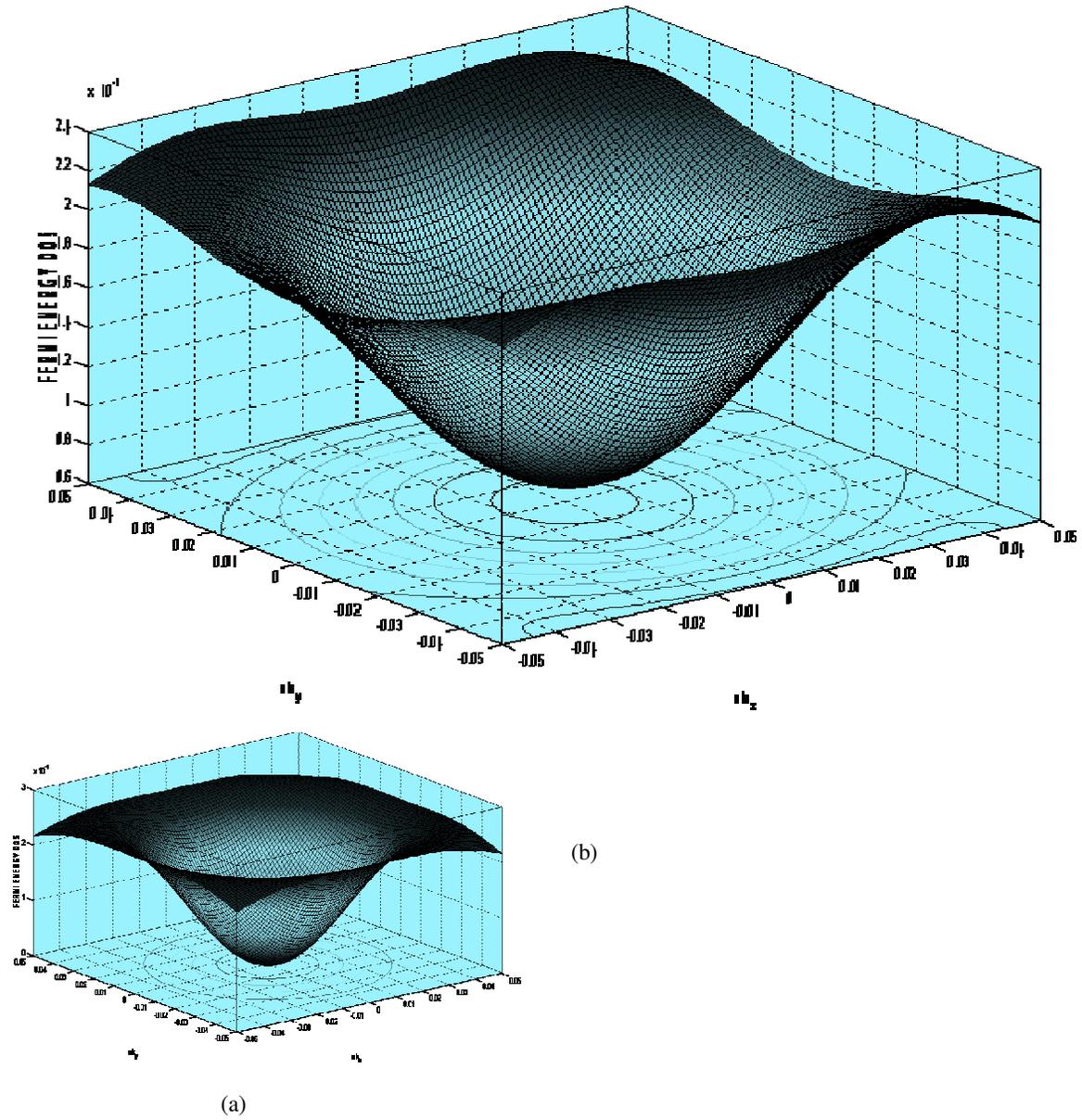

**Figure 4 .** The 3D plots of the Fermi energy density of states corresponding to the AA-stacking, in the momentum space, for (V/ $\gamma_1$) = 0.0 (Figure(a)), and (V/ $\gamma_1$) = 0.17 (Figure(b)) . The inverted sombrero-like structure in (a) gets deformed due to the increase in bias.